\documentclass[reprint,aps,pra,a4]{revtex4-2}
\usepackage[T1]{fontenc}
\usepackage[latin9]{inputenc}
\usepackage{verbatim}
\usepackage{amssymb}
\usepackage{graphicx}
\usepackage{amsmath}
\usepackage[bookmarks=false,linkcolor=blue,urlcolor=blue,colorlinks,citecolor=blue]{hyperref}
\usepackage{xcolor}
\usepackage{babel}

\begin{document}

\title{Engineered platforms for topological superconductivity and Majorana zero modes}

\author{Karsten Flensberg}
\affiliation{\mbox{Center for Quantum Devices, Niels Bohr Institute, University of Copenhagen, DK-2100 Copenhagen, Denmark}}

\author{Felix von Oppen}
\affiliation{\mbox{Dahlem Center for Complex Quantum Systems and Fachbereich Physik, Freie Universit\"at Berlin, 14195 Berlin, Germany}}

\author{Ady Stern}
\affiliation{Department of Condensed Matter Physics, The Weizmann Institute of Science, Rehovot 76100, Israel}

\begin{abstract}
Among the major approaches that are being pursued for realizing quantum bits, the Majorana-based platform has been the most recent to be launched. It attempts to realize qubits which store quantum information in a topologically-protected manner. The quantum information is protected by its nonlocal storage in localized and well-separated Majorana zero modes, and manipulated by exploiting their nonabelian quantum exchange properties. Realizing these topological qubits is experimentally challenging, requiring superconductivity, helical electrons (created by spin-orbit coupling) and breaking of time reversal symmetry to all cooperate in an uncomfortable alliance. Over the past decade, several candidate material systems for realizing Majorana-based topological qubits have been explored, and there is accumulating, though still debated, evidence that zero modes are indeed being realized. This paper reviews the basic physical principles on which these approaches are based, the material systems that are being developed, and the current state of the field. We highlight both the progress made and the challenges that still need to be overcome.
\end{abstract}

\maketitle

\section{Introduction}

In the realm of condensed matter physics, Majorana zero modes (MZMs)  \cite{Kitaev2001,Read2000} are supposed to be the cake we can both eat and have. The cherries in the cake are the qubits of quantum information that the MZMs  are supposed to host. Having the cake here means having qubits that, due to their topological protection, are not easily measured or disturbed by their environment. These qubits are expected to enjoy very long coherence times, growing exponentially with the distance between the Majoranas and the inverse of the temperature.  On the other hand, eating the cake means interacting with the qubits, manipulating and reading their states. This is enabled by well-designed ways to measure and affect the state of the qubits, ways that are accessible to us, but not to a randomly fluctuating environment \cite{Kitaev2003,NayakReview}.

The beauty of this vision is obvious to anyone who has ever experienced the desire to eat a cake and have it, too. This short review is aimed at giving the reader a status report on the baking of this cake. We introduce the general concepts underlying the field, survey several material systems that are being experimentally studied, discuss the difficulties in identifying the MZMs in these experiments, and conclude with an outlook for the future. The research effort involved is rather broad in scope, and is difficult to cover in a short review paper. Various aspects of the field have already been described in several review papers \cite{AliceaReview,BeenakkerReview,LutchynReview,Aguado2017,Haim2019,RamonLeoPhysicsToday,OregOppenAnnualReview,Prada2019}. The angle we chose here is to cover broadly the experimental systems that are being explored in attempts to discover zero-dimensional localized Majorana zero modes (MZMs) in engineered systems combining metals, semiconductors, or superconductors. Among other things, our choice leaves out theoretical models with no immediate candidates for realization and candidate systems for realizing one-dimensional Majorana modes.

The key concept behind the notion of Majorana fermions, or MZMs, as they are more properly called in the condensed matter context, is that of a topologically-protected degeneracy of the ground state. In a system with $2N$ MZMs, the ground state is $2^N$-fold degenerate. The MZMs are described by operators $\gamma_i$, with $i=1,\ldots, 2N$, that take the system from one ground state to another, at zero energy cost, hence the name zero modes. These operators are hermitian and obey fermionic commutation rules, and hence the name Majorana fermions (see Box 1 for more information). The number of such modes must be even, and it is this constraint that protects them. If the MZMs are localized and spatially distant from one another, a local perturbation cannot shift any of them away from zero energy. As a consequence, both the ground-state degeneracy and the state vector in this manifold are protected against any local perturbation, or at least any perturbation that does not add or remove single electrons from the system. Single-electron processes, however, are exponentially suppressed due to the existence of an energy gap.

For a fermionic operator to be hermitian, it must be a superposition of a creation and an annihilation operator, and such excitations occur primarily when superconductors are involved. Thus, most realizations of Majorana fermions involve a superconducting element in the system. Conceptually the most straightforward of all, the Kitaev chain, is a one-dimensional system of spinless fermions which are made superconducting by a $p$--wave order parameter. In the topological phase, the chain hosts one MZM at each of the two ends \cite{Kitaev2001}. Similarly, in a two-dimensional $p$--wave topological superconductor, the core of a vortex hosts a MZM \cite{Kopnin1991}.

Unfortunately, conceptual simplicity and natural abundance do not go hand in hand here, and $p$--wave superconductors are, at best, rare. Fortunately, their topological properties can be realized based on more conventional $s$--wave superconductors, if the latter are combined with spin-orbit coupling and a magnetic field (or another time-reversal-symmetry breaking agent). Indeed, most candidate systems for realizing and exploring MZMs involve these three ingredients - spin-orbit coupling, magnetic field, and superconductivity. Below, we survey different ways in which these ingredients are combined (Sec.\ \ref{Sec:realizations}), before discussing the extent to which MZM physics has already been realized (Sec.\ \ref{sec:prove_disprove}) and providing an outlook (Sec.\ \ref{Sec:outlook}).

\section{Platforms and observations}
\label{Sec:realizations}

\subsection{Topological insulators}
\label{sec:FuKane}

Majorana zero modes would be much more commonplace if only they were carried by the ``usual suspects'' of superconductivity -- $s$-wave superconductors such as aluminum or niobium. The reason this is not the case is that a MZM is an isolated zero mode which is localized far from any other zero mode. For $s$-wave superconductors, the symmetry of the pairing makes each excitation mode doubly degenerate, a degeneracy whose roots lie in the Fermi surface having two degenerate copies, one for each spin state as dictated by Kramers' theorem.

With the existence of two Fermi surfaces being an obstacle to realizing MZMs, there are two directions to pursue for reducing that number - either looking for systems which have only a single Fermi surface, or starting with a system that has two surfaces and finding a mechanism to gap one of them. The first direction was suggested by Fu and Kane \cite{Fu2008a,Fu2008b}, building on two observations. First, the one-dimensional edge state of a two-dimensional topological insulator has a single Fermi surface, here in the form of a single pair of Fermi points with opposite spin directions \cite{Bernevig2006,Fu2006}. And second, the two-dimensional surface of a three-dimensional topological insulator hosts a single Dirac cone, in which the spin direction is locked to the direction of the momentum \cite{Fu2007topological,Fu2007,Moore2007topological,Qi2008topological}. Both systems have only a single Fermi surface, but still allow for efficient coupling to an $s$-wave superconductor through the proximity effect as electronic states of opposite momenta carry opposite spins.

The one-dimensional edge state of a two-dimensional topological insulator may be gapped in two ways. A gap may develop by spin-flip scattering between opposite momenta, which is allowed only when time-reversal symmetry is broken. Alternatively, superconductivity gaps the edge of the two-dimensional topological insulator by forming a Cooper pair with the two electrons being from the two counter-propagating branches, breaking charge conservation on the edge.  When adjacent regions of the edge are gapped by these two different mechanisms, the transition from one type of gapping to the other requires a closure of the energy gap, and the localized gapless mode at the interface is a MZM.

In the case of a three-dimensional topological insulator, proximity to an $s$-wave  superconductor creates a superconductor of a single Fermi surface, but not yet a MZM, for two reasons. First, we need to find a location for the MZM to localize. And second, we need to break time reversal symmetry that enforces a twofold Kramers degeneracy of the excitation modes. Both needs are addressed by introducing vortices in the proximity-coupled superconductor. Indeed, the core of each such vortex carries a single MZM.

The observation of vortex-bound MZMs is complicated by the existence of additional (Caroli-Matricon-de Gennes) states localized in the vortex cores. For weak-coupling superconductors, with the energy gap $\Delta$ much smaller than the Fermi energy $E_F$, the energy of these additional intra-core states can be close to zero, with a level spacing of $\Delta^2/E_F$. Thus, the spectral difference between topological and trivial superconductors is that the lowest state has zero energy for the former and an energy of order $\Delta^2/2E_F$ for the latter. Since $\Delta^2/2E_F$ is typically much smaller than attainable temperatures, this difference is challenging to detect. Other differences involve the shape and spin polarization of the wave function of the intra-core states \cite{HeLeeLaw}, their magnetic field dependence, and the difference between MZMs carried by vortices and anti-vortices \cite{KrausAuerbach}.

Experimentally, a natural tool to observe a MZM in a vortex core is a scanning tunneling microscope (STM). However, if the MZM is buried at the interface between a topological insulator and a superconductor, the STM cannot approach it. An elegant solution to this inherent problem may be provided by iron-based supercondutors, in which several bands overlap in energy, with one band being topological and insulating, hence carrying a surface state, and the other bands forming a 3D superconductor that proximity couples to the surface state. Experimental studies of these surfaces have discovered states at the cores of vortices \cite{Wang333}, but their identification as MZMs is still under debate.

Alternatively, MZMs can be created on a topological-insulator surface along a one-dimensional domain wall separating an island which is gapped, say, by an $s$-wave superconductor from its surrounding, which is gapped by a magnetic field (or the exchange field of a ferromagnet). In this case, the MZM is a one-dimensional object and part of a propagating chiral Majorana mode, whose direction of propagation can be reversed by switching the direction of the magnetization, or the roles of superconductor and magnet  \cite{Fu2008a}.

\subsection{Engineered semiconductor-superconductor platforms}

Inspired by the Fu-Kane proposal to use topological insulators and their built-in nontrivial topological structure, researchers realized that a similar situation can be engineered by a combination of more conventional materials with sufficiently strong spin-orbit coupling, proper breaking of time-reversal symmetry and, as in the Fu-Kane scheme, a strong interface to an ordinary  $s$-wave superconductor \cite{Sau2010,Alicea2010,Lutchyn2010,Oreg2010,Chung2011,Duckheim2011,Potter2012}. The time-reversal breaking agent could be the exchange field of a ferromagnetic insulator \cite{Sau2010}, layers of half-metals  \cite{Duckheim2011,Chung2011}, or simply an externally applied magnetic field \cite{Lutchyn2010,Oreg2010}.

Several of these proposals have since been experimentally realized
 \cite{Mourik2012,Krogstrup2015,Chang2015,Gul2018,Zhang2017,Liu2019b} and characterized by transport measurements using nanofabricated devices \cite{Mourik2012,Deng2012,Rokhinson2012,Churchill2013,Das2012,Chang2015,Higginbotham2015,Albrecht2016,Deng2016,Suominen2017,Chen2017,Gul2018,Chen2017,Nichele2017,Sestoft2018,Vaitiekenas2020,Vaitiekenas2020b}. Below, we review the various platforms and discuss their advantages and disadvantages.

\subsubsection{Semiconductor quantum wires}

As mentioned in Sec.\ \ref{sec:FuKane}, creating MZMs requires a single -- or more generally an odd number of -- Fermi surfaces. For a quasi-one-dimensional material, this corresponds to an odd number of Fermi points with positive momenta. This can, of course, be arranged by applying a sufficiently strong magnetic field to a wire and adjusting the chemical potential such that, say, only the lowest spin-split energy band is occupied. However, gapping such a spin-polarized system by superconducting pairing would seem to require a triplet superconductor. Extending ideas developed for two-dimensional systems \cite{Fu2008a,Sau2010,Alicea2010} (see also below), it was proposed by Lutchyn \textit{et al.} \cite{Lutchyn2010} and Oreg \textit{et al.} \cite{Oreg2010} that the combination of Rashba spin-orbit coupling and conventional $s$-wave superconductivity removes the necessity for intrinsic triplet pairing.

The Rashba spin-orbit coupling exposes the electron spin to an effective Zeeman field whose direction is perpendicular both to  the electric field exerted on the electron by the crystal and to the electron's velocity.  This effective Zeeman field points in opposite directions for electrons of opposite velocities, so that the spin-orbit coupling twists the spin directions at the two Fermi points in different directions. As a consequence, a pair of electrons with opposite momenta now has both a triplet and a singlet component, allowing them to be proximitized by a conventional superconductor. When the electron dispersion is approximated as parabolic, the corresponding energy scale is parametrized as $\alpha k_F$, where $\alpha$ encodes the magnitude of the electric field and $k_F$ is proportional to the velocity.

For a strong magnetic field, $B$, and weak spin-orbit coupling, $\alpha k_F$, this physical picture implies that the proximity-induced gap in the nondegenerate band (i.e., the ``topological gap'') is proportional to the product of the gap $\Delta_\mathrm{ind}$ induced by the superconductor into the semiconductor and the angle by which spin-orbit coupling tilts the spins away from being parallel to one another, $\alpha k_F /g\mu_B B$. (Here $g$ is the semiconductor's $g$-factor, and $\mu_B$ is the Bohr magneton.) It is this tilting that enables proximity coupling between a semiconductor and an $s$-wave superconductor, where Cooper pairs are spin singlets. The important material parameters are thus the spin-orbit coupling and the $g$-factor. From the above estimate, it would seem that the $g$-factor should be small such that the gap becomes large, but this in fact is not the case, because the Zeeman energy, $\frac{1}{2}g\mu_B B$, should also be larger than the induced gap, for the system to enter the topological phase. In fact, the precise criterion is $\frac{1}{2}g\mu_B B > \sqrt{\Delta_\mathrm{ind}^2+\mu^2}$ \cite{Lutchyn2010,Oreg2010}, where $\mu$ is the chemical potential of the unsplit one-dimensional band, measured with respect to the midpoint of the Zeeman-induced gap. From this expression, we see that the Zeeman energy needs to be larger than the induced gap. At the same time, the Zeeman energy should not suppress the superconducting gap $\Delta_S$ of the parent superconductor too strongly, and one therefore also requires $g_S \mu_B B < \Delta_S$ (with $g_S$ the $g$-factor of the superconductor). The optimal parameter regime is thus large $g/g_S$, large $\alpha$, and large $\Delta_\mathrm{ind}/\Delta_S$.

\begin{figure*}
\centerline{\includegraphics[width=\textwidth]{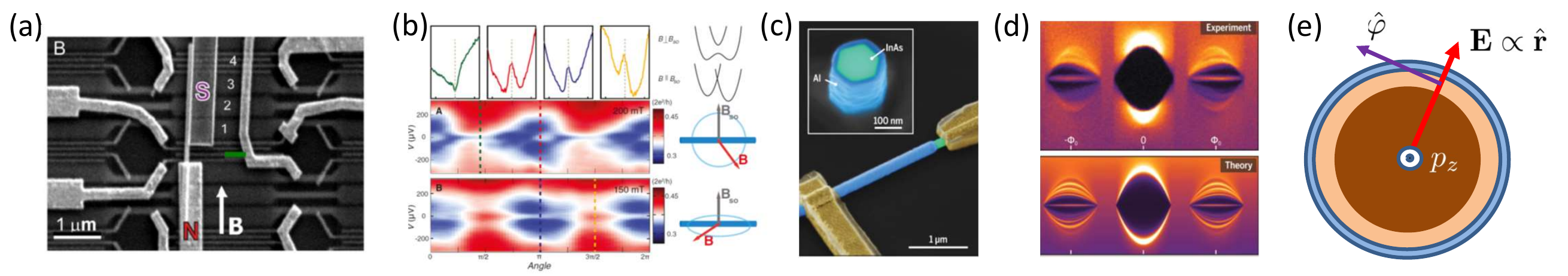}}
\caption{\label{fig:wires} Two realizations of engineered topological superconductors based on semiconducting nanowires proximitized by conventional superconductors. The first experiment \cite{Mourik2012} from Delft is shown in panels (a) and (b). (a) shows the device layout where a nanowire is contacted by a superconductor (labelled S) which is evaporated onto one side of the nanowire. Also shown are gates to control the density in the wire (labelled 1,2,3,4) and the tunnel barrier (green marker) to the normal probe (labelled N). In (b), the differential conductance is shown for different angles of the applied magnetic field corresponding to the colors in the panels below. Consistent with the expectation, the zero-bias peak is most prominent for $B$ perpendicular to the expected spin-orbit field $B_\mathrm{SO}$. The band-structure sketches to the right illustrate the difference between the two $B$-field directions. Only for $B\perp B_\mathrm{SO}$ does the system have a single Fermi point for positive wave numbers. An alternative approach to induce a topological state is shown in panels (c)-(e). Here the wire is fully covered by the superconductor, see panel (c) for a cross section and the device geometry. A topological state can be realized when the shell is pierced by more than half a flux quantum such that the energy is minimized when the phase of superconducting order parameter in the shell has a winding number $n=1$. In this situation, a zero-bias peak is indeed observed, in agreement with a detailed theoretical calculation (panel d). This is made possible by a spin-orbit field in the azimuthal direction which defines an angular direction (e) relative to the phase winding and can thus cause the system to become helical.
Panels reprinted from Refs.: (a) and (b) \cite{Mourik2012}; (c) and (d) \cite{Vaitiekenas2020}.
}
\end{figure*}

In the original theory papers \cite{Lutchyn2010,Oreg2010}, it was suggested that this rather precise set of requirements can be met in InAs or InSb nanowires in proximity to a conventional superconductor with not too small a pairing gap. Indeed, the materials of choice of the first experiment \cite{Mourik2012} were InSb nanowires and a NbTiN superconductor which was evaporated to cover part of the wire. Using the sample shown in Fig.\ \ref{fig:wires}a, the experiment probed the subgap spectrum at the end of the proximitized nanowire by tunneling spectroscopy from a normal metal (Au) lead (see Box 2 for further discussion of  spectroscopic signatures of MZMs). The results were exciting and promising because they were consistent with an interpretation in terms of Majorana end states. These observations included the appearance of a zero-bias peak at a finite magnetic field, which disappeared again at larger fields. This is consistent with the magnetic field first driving the system from a trivial phase at low fields to a topologically nontrivial phase at intermediate fields and then inducing a reemergence of a trivial phase at yet higher fields. Moreover, theory predicts a distinct dependence on the magnetic-field direction. The optimal field orientation for inducing a topological state is perpendicular to the spin-orbit field, so that the latter maximally twists the electron spins. Since the spin-orbit field points perpendicular to both the propagation direction of the electrons and the interface between superconductor and semiconductor, the plane of optimal magnetic-field orientations can be deduced from the device layout. Mourik \textit{et al.} \cite{Mourik2012} probed this prediction by tilting the magnetic field out of this plane and observed, consistent with the Majorana picture, that the peak and the ``topological gap'' vanish when the field aligns with the presumed direction of the spin-orbit field, see Fig.\ \ref{fig:wires}b.

Encouraged by the initial experiments \cite{Mourik2012,Deng2012,Rokhinson2012,Das2012,Churchill2013}, other labs picked up the challenge of producing better hybrid materials. A major breakthrough came with the invention of epitaxially grown hybrid semiconductor-superconductor systems \cite{Krogstrup2015,Chang2015,Sestoft2018} in which some facets of the semiconducting wires are covered by a superconducting shell. The cleaner and more homogenous interface was shown to be key to induce a hard gap in the semiconductor. This type of devices led to further studies of near zero-bias peaks, as well as more detailed topological phase diagrams \cite{Deng2016,Chen2017,Deng2018,Gul2018}. According to theory, the two Majorana bound states at the ends of the wire split away from zero energy, when they overlap. As a consequence, one should also observe two conductance peaks. Tunneling spectroscopy with a quantum dot between the normal lead and the topological superconductor provides a means to study this splitting with improved sensitivity. Corresponding experiments were reported in Ref.\  \cite{Deng2016}, later refined both experimentally \cite{Deng2018} and theoretically \cite{Prada2017,Clarke2017}.

A key prediction for the Majorana zero-bias peak of a normal metal-topological superconductor junction is that its height is quantized to a value of $2e^2/h$ \cite{Sengupta2001,Law2009,Flensberg2010,Wimmer2011}. This is, however, not what is observed in experiments. One complication is that a zero-bias peak at the quantized value is only reached when temperature is much lower than the broadening of the Majorana resonance induced by the tunnel coupling between topological superconductor and normal lead. At the same time, the broadening should be much smaller than the topological gap, typically leaving a rather small window. One study showed that by correlating the tunnel coupling (as determined from the above-gap conductance) and the width (as determined by fitting to a thermally broadened Lorentzian-shaped Majorana peak), one could obtain scaling behavior consistent with a zero-temperature quantized conductance \cite{Nichele2017}. As a word of caution, it should be noted that in certain cases, a conductance quantized at $2e^2/h$ can also result from trivial states \cite{Yu2020,pan2021,zhang2021}.

Tunneling spectroscopy as discussed so far can only give indirect information about the nonlocal properties of Majorana bound states. Other transport measurements which provide somewhat complementary information include Coulomb-blockade  \cite{Albrecht2016,Albrecht2017,Shen2018} or three-terminal experiments \cite{Menard2019}. Coulomb-blockade experiments \cite{Albrecht2016} using finite segments of superconducting islands found a transition from 2$e$-periodic to $1e$-periodic Coulomb-blockade peaks, as expected from Majorana physics \cite{Fu2010,Heck2016}. Moreover, for finite-length wires the coupling between the Majoranas would predict an alternating energy difference for even and odd island occupations. This was indeed observed and the extracted splitting was found to be consistent with the expected exponential decay as a function of wire length. More details on Coulomb blockade signatures of MZMs are given in Box 2.

Proximitized nanowires are presumably the best-studied Majorana-candidate system, with numerous detailed experiments -- including tunneling spectroscopy, Coulomb-blockade measurements, multi-terminal transport, or Josephson junctions -- probing various aspects of the expected Majorana physics. While several observations are consistent with a Majorana interpretation, it should be noted that the observed signatures may also have alternative interpretations. This is discussed further in Sec.\ \ref{sec:prove_disprove} below.

One challenging aspect of the proximitized nanowires discussed so far is that the magnetic field required to enter the topological regime is frequently rather high. This may induce states within the superconducting gap and soften the gap considerably. The required spin splitting can alternatively be induced by proximity to a ferromagnetic insulator, as was proposed in a variety of contexts \cite{Fu2008a, Fu2008b, Sau2010,Alicea2010}.
It recently became possible to grow the ferromagnetic insulator EuS on InAs \cite{Liu2019a}, producing nanowires with a hexagonal cross section, of which three facets each are covered by the superconductor and the ferromagnetic insulator \cite{Liu2019b,Vaitiekenas2020b}. When one facet has overlapping covers, it appears that a transition to a state with a zero-bias peak is induced \cite{Vaitiekenas2020b}. The physics of this transition is not yet fully understood and calls for further investigation.

\paragraph{Full-shell quantum wires.}

Wires which are entirely surrounded by a layer of superconductor are referred to as full-shell wires (Fig.\ \ref{fig:wires}c and e). This geometry clearly has a different symmetry from the partially-covered wires. Instead of the Rashba spin-orbit coupling creating an effective Zeeman field in a fixed direction perpendicular to the wire, here the effective Zeeman field  must by symmetry point in the azimuthal  direction. Naively, one might therefore guess that its net effect averages out.

Despite this intuition, an experimental group went ahead and investigated the end-state spectrum of corresponding nanowires by tunneling spectroscopy \cite{Vaitiekenas2020}, finding a zero-bias peak when the magnetic flux through the wire causes the superconducting phase
to wind by 2$\pi$ around the aluminum shell and thus creates a fluxoid in the induced pairing (see Fig.\ \ref{fig:wires}d). The change in winding number,  related to the Little-Parks effect \cite{Tinkham2004}, was indeed key to understanding the possible transition into a topologically nontrivial state. Although the effective Zeeman field is azimuthal, it introduces a handedness to the problem, when combined with the phase winding of the superconducting phase: For electrons moving in one direction, the spin-orbit field is along the phase winding, whereas for electrons moving in the opposite direction, the spin-orbit field winds against the superconducting phase. Based on this observation, Ref.\ \cite{Vaitiekenas2020} developed a Majorana-based theory and numerical modelling that agreed with the data to some extent. This interpretation was challenged by a later experiment \cite{Valentini2020}, that interpreted the zero-bias peak as originating from Andreev states. Full-shell wires enjoy the advantage that the magnetic field needed to induce a phase winding is much smaller than the critical Zeeman field of partially covered wires. A disadvantage is that full-shell coverage makes it impossible to tune the electron density by gating, eliminating an important control for tuning the wire into the topological phase.

\subsubsection{Two-dimensional platforms.}

\begin{figure*}
\centerline{\includegraphics[width=1.\textwidth]{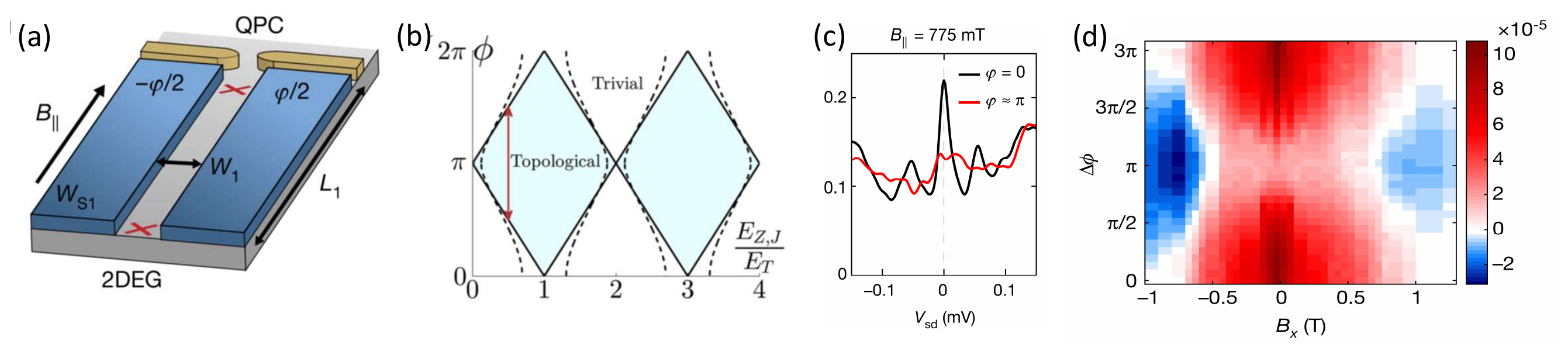}}
\caption{Realization of one-dimensional topological superconductivity in a superconductor-normal-superconductor Josephson junction ($N$-stripe). (a) The setup. A superconductor-normal-superconductor Josephson junction, where the normal part is subject to Rashba spin-orbit coupling, the two supeconductors are maintained at a phase difference $\phi$, and the system is exposed to a parallel magnetic field $B_\parallel$ coupling to the electrons' spins. (b) Phase diagram as a function of $B_{\parallel}$ and $\phi$. (c) Differential conductance measured by tunneling from an external normal lead via the quantum point contact (QPC) shown in (a). A zero-bias peak emerges when the phase difference is zero, and is absent when the phase difference is $\pi$. (d) The sign of the second derivative of the differential conductance, indicating the presence of a zero-bias peak, as a function of $B_\parallel$ and $\phi$. Panels reprinted from Refs.: (a) and (c) \cite{Fornieri2019}; (b) \cite{Pientka2017}; (d) \cite{Ren2019}.
\label{fig:jj}}
\end{figure*}

The same technique that allows near-perfect epitaxial growth of thin aluminum layers on GaAs or InAs wires, can also be used to grow two-dimensional layers of superconductor on a semiconductor heterostructure, based for instance on InAs \cite{Shabani2016}. This allows for new and interesting design possibilities of one-dimensional topological superconductors. The first
One geometry, referred to as $S$-stripes, is formed by removing the superconducting cover everywhere except along a narrow one-dimensional strip and depleting the exposed electron gas by a negatively charged gate \cite{Hell2017}. In this way, one effectively creates a situation similar to the partially-covered nanowires discussed earlier. A second option, referred to as $N$-stripes \cite{Hell2017,Pientka2017} and depicted in Fig.\ \ref{fig:jj}a, has in some sense an inverted geometry. The superconductor is removed only along a narrow stripe, leaving an exposed semiconductor not covered by the superconductor. This effectively realizes  an ``Andreev quantum well'' or a wide Josephson junction, which binds one-dimensional electronic states within the energy gap of the superconductor. Both geometries were studied theoretically as well as experimentally, as we review in the following.

\paragraph{One-dimensional wires in a two-dimensional setting ($S$-stripes).}

The $S$-stripe geometry showed zero-bias peaks similar to what was seen in nanowires. This was used to perform the above-mentioned study of the temperature dependence of the peaks for various tunnel couplings \cite{Nichele2017}, which suggests that the conductance measurements are consistent with a zero-bias conductance of $2e^2/h$ at low temperatures. The new capability of designing more complicated networks provided by the two-dimensional platform was exploited to realize an interferometer with a Majorana wire incorporated into one its arms \cite{Whiticar2019}.

\paragraph{Planar Josephson junctions ($N$-stripes).}

The quantum-wire realization of one-dimensional topological superconductivity exploits the chemical potential, the Zeeman field, and the superconducting pair potential as tuning parameters. Within this parameter space, there are regions where the wires carry Majorana end modes. In particular, the chemical potential needs to be tuned to a region in which one pair of Fermi points is gapped by the Zeeman field, while the other is gapped by superconductivity. Tuning the chemical potential is experimentally challenging, given that the potential exerted by metallic gates is heavily screened by the superconductors in proximity to the wires.

The Josephson junction geometry ($N$-stripes) considerably weakens the dependence of the phase diagram on the chemical potential and also reduces the magnetic field required for achieving topological superconductivity \cite{Hell2017,Pientka2017}. In this setup, depicted in Fig.\ \ref{fig:jj}a, a 2DEG with strong spin-orbit coupling replaces the quantum wire, while the superconductor that couples to the wire is replaced by superconductors on both sides of the junction. The phase difference between these superconductors constitutes a useful additional tuning parameter. A Zeeman field pointing parallel to the normal channel is required here, as well. The topological superconductivity is realized in the quasi-one-dimensional channel between the two superconductors, and transitions from topological to trivial phases take place through a closure of the energy gap within the channel.

In the ideal limit of semi-infinite superconducting banks and spin-orbit coupling much stronger than the Zeeman energy, the phase diagram depends primarily on two parameters, the phase difference $\phi$ between the two superconductors and $E_Z W/v_F$, the Zeeman energy multiplied by the time it takes an electron to cross the junction of width $W$. (Here, $v_F$ is the Fermi velocity). The phase diagram is shown in Fig.\ \ref{fig:jj}b, and shows topological superconductivity in half of the parameter plane.  In particular, when $\phi=\pi$, an arbitrarily weak Zeeman energy is in principle sufficient to drive a clean Josephson junction into the topological phase \cite{Hell2017,Pientka2017}.

Although the topological superconductivity realized in this system is one dimensional, the system itself is two dimensional, since the normal part may in principle host many subbands. Semiclassically, then, the electron motion has components both along and across the junction and the magnitude of the velocity is equal to the Fermi velocity. Every reflection from one of the superconductors involves an amplitude for Andreev reflection, converting an electron to a hole or vice versa, as well as an amplitude for a normal reflection, in which the reflected particle is not converted. In the ideal limit as defined above, the Andreev amplitude dominates. Away from the ideal limit,
the phase dependence weakens, the dependence on chemical potential sharpens, and the junction becomes more similar to the quantum-wire case \cite{Setiawan2019}.

For an infinitely long junction, the spectrum of in-gap excitations associated with the normal channel may be characterized by the wavevector $k_x$ along the channel. The topological phase diagram can be deduced from the properties of the junction at $k_x=0$, which  corresponds semiclassically to electrons and holes propagating across the junction. The energy gap, in contrast, is smallest for electrons that move primarily along the channel, since they are least affected by the superconducting banks. Realizing and probing a topological phase in experiment frequently depends on a sufficiently large energy gap. Here, enhancing the gap requires means to limit trajectories that move long distances parallel to the superconducting banks.

A simple and straight-forward way to limit such a trajectory is to design the normal part of the junction in a zig-zag shape, forcing the semiclassical trajectories to undergo Andreev reflections whenever the zig-zag changes direction \cite{ZigzagAkhmerov}. This approach comes at a price, however. In a zig-zag geometry, a uniform magnetic field inevitably has a component perpendicular to the local direction of the channel. This tends to suppress the energy gap, and the details of the zig-zag pattern need to be optimized to maximize the gap. A possible approach to this optimization replaces the Zeeman field by an applied supercurrent along the junction \cite{Akhmerovcurrent}.

Long parallel trajectories are also naturally limited by disorder in the channel. Indeed, when the effect of weak disorder is analyzed perturbatively, it is found to enhance the localization of the MZMs close to the junction ends \cite{ArbelDisorder}.

This setup has been examined experimentally by several groups, using different material systems. In one experiment \cite{Ren2019}, see Fig.\ \ref{fig:jj}d, the 2DEG was based on a HgTe quantum well, coupled to superconducting Al leads. In two  experiments \cite{Fornieri2019,Shabani2019}, see Fig.\ \ref{fig:jj}c, the 2DEG was formed in an InAs heterostructure, with the superconductivity again provided by aluminum. And a fourth experiment used an InSb-based 2DEG and NbN$_3$ as the superconductor \cite{Goswami2019}. Two of these experiments \cite{Ren2019,Fornieri2019} measured the tunneling density of states of the junction near its ends, where MZMs are expected to appear in the topological phase, and observed zero-bias peaks for parameter values, which are consistent with the expected phase diagram.

\subsection{Chains of magnetic adatoms}

\begin{figure*}
\centerline{\includegraphics[width=1.\textwidth]{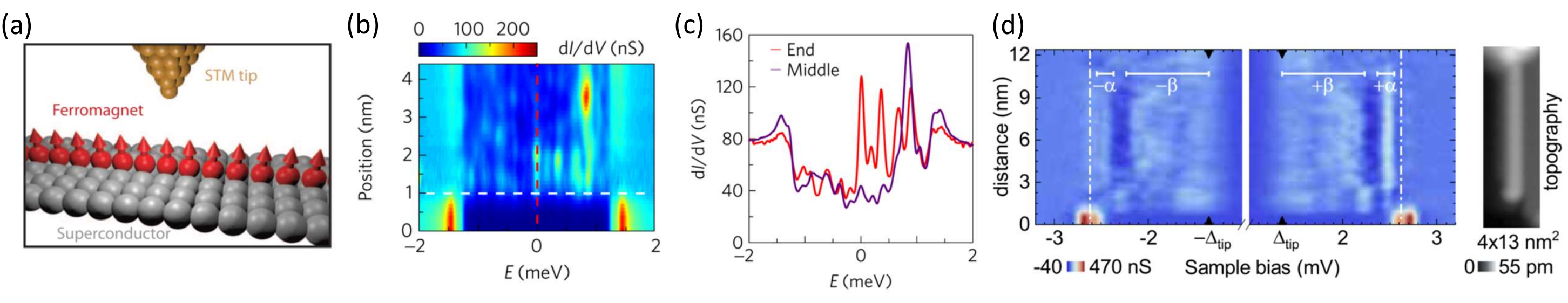}}
\caption{\label{fig:chains}
Topological superconductivity and Majorana zero modes in chains of magnetic adatoms on superconductors. (a) Schematic rendering of a chain of magnetic adatoms on a superconductor, probed by an STM. (b) High-resolution $\mathrm{d}I/\mathrm{d}V$ (color scale) spectroscopy along a chain of Fe adatoms on a Pb(110) substrate, using a normal-state STM tip, showing a zero-bias peak localized within 1nm of the end of the wire (indicated by dashed white line). (c) Representative line cuts show a pronounced zero-bias peak at the end of the wire, but no such peak away from the end. (d) Corresponding data on a Co chain, showing no signatures of Majorana zero modes (now expected at voltages equal to the tip gap, $eV=\pm \Delta_{\rm tip}$ due the use of a superconducting tip for improved resolution. Panels reprinted from Refs.: (a) \cite{NadjPerge2014}; (b) and (c) \cite{Feldman2017}; (d) \cite{Ruby2017}. }
\end{figure*}

A paradigmatic model for one-dimensional topological superconductors is provided by the Kitaev chain with nearest-neighbor hopping and pairing between adjacent single-orbital sites \cite{Kitaev2001}. A rather faithful realization of this model can be imagined in chains of magnetic adatoms on conventional superconducting substrates \cite{NadjPerge2013,Pientka2013}, see Fig.\ \ref{fig:chains}a. An individual magnetic adatom induces a pair of spin-polarized subgap  states \cite{Yu1965,Shiba1968,Rusinov1969} in the substrate superconductor. These states are referred to as Yu-Shiba-Rusinov (YSR) states. When adatoms are in sufficiently close proximity, their YSR states hybridize with one another. In a chain of such YSR states, hopping between neighboring sites broadens the particle-hole symmetric pair of YSR states into bands. Once the bands formed by the negative and positive-energy YSR states overlap at the center of the gap, $p$-wave pairing correlations can reopen a gap and turn the adatom chain into a one-dimensional topological superconductor with Majorana end states.

The mechanism of inducing $p$-wave pairing depends on the magnetic ordering of the adatoms. The YSR states are spin polarized, so that the spin-singlet Cooper pairs of $s$-wave superconductors cannot proximity couple to a strictly ferromagnetic chain. Proximity coupling becomes possible in the presence of spin-orbit coupling, for instance in the substrate superconductor. Alternatively, $p$-wave superconducting correlations can be induced when the adatoms form a helical magnetic state in which the spin polarization of the YSR states rotates along the chain, thus creating an alternative coupling between spin and orbital degrees of freedom \cite{Braunecker2010,Kjaergaard2012}.

Several works suggested that a topological phase can form in a self-organized fashion without a need for tuning the chemical potential \cite{Braunecker2013,Klinovaja2013,Vazifeh2013,Shecter2016}. Coupling a one-dimensional electronic system to magnetic moments induces a tendency to form a spiral spin structure as a result of the Ruderman-Kittel-Kasuya-Yosida (RKKY) interaction. Due to its wavevector of $2k_F$ which is inherently tied to the Fermi wavevector $k_F$ of the electronic channel, the spin spiral opens a partial gap at the Fermi level. In conjunction with superconducting correlations, this leads to a topological superconducting phase without tuning of the chemical potential.

A realization of these dilute-chain scenarios for topological superconductivity remains an outstanding experimental challenge. Existing experiments on topological superconductivity with magnetic adatoms are presumably based on chains in which the adatoms themselves are strongly coupled. Direct hybridization of the adatom $d$ orbitals causes the formation of spin-split bands \cite{Li2014}. With the same ingredients as for dilute chains, the substrate superconductor can then induce $p$-wave correlations in those bands that cross its Fermi energy. Topological superconductivity ensues for an odd number of partially filled bands.

First experimental evidence for MZMs in chains of magnetic adatoms was found by Nadj-Perge {\em et al.} \cite{NadjPerge2014} using a self-organized chain of Fe adatoms on a superconducting Pb substrate. Probing the chains by scanning tunneling spectroscopy (STS) using a normal-metal STM tip revealed zero-bias peaks localized at the end of the chain, but no such peaks elsewhere along the chain (see Fig.\ \ref{fig:chains}b and c for low-temperature data). Measurements with a spin-polarized tip suggest that the adatoms order ferromagnetically. Surprisingly, the end states were strongly localized, on a scale much shorter than a naive estimate of the coherence length of the induced superconductivity. This has been interpreted as a result of renormalizations resulting from the strong coupling of the adatom $d$ orbitals to the substrate superconductor \cite{Peng2015}, which can also be understood from a semiclassical perspective \cite{Kiendl2019}.

Measurements with superconducting tips can improve resolution owing to the suppression of thermal excitations by the superconducting gap and the sharp coherence peak in the superconducting density of states. With superconducting tips, MZMs no longer appear as zero-bias peaks. Since the tunneling process leaves an unpaired electron behind in the tip, they rather appear at a bias voltage given by the superconducting gap of the tip (see also Box 2). The particle-hole symmetry of the MZMs implies that there must be symmetric peaks at positive and negative bias voltages \cite{Ruby2015}. Experimental results probing this symmetry are promising, but not yet fully conclusive \cite{Ruby2015,Feldman2017}. With superconducting tips, identification of zero-energy states generally requires accurate knowledge of the tip gap. The energy of subgap states can, however, be measured directly as a line splitting in photon-assisted tunneling experiments \cite{Peters2020}.

Additional experiments with Fe chains on Pb substrates have probed the zero-energy end states by spin-polarized tips \cite{Jeon2017,Li2018} or atomic force microscopy \cite{Pawlak2016}. Experiments using Co chains on a Pb substrate did not exhibit signatures of MZMs \cite{Ruby2017}, see Fig.\ \ref{fig:chains}d. As Co has an additional $d$ electron compared to Fe, this has been interpreted as reflecting an even number of occupied $d$ bands. Possible evidence for topological superconductivity and MZMs has, however, been observed in two-dimensional islands of Co on Si, covered by a monolayer of Pb \cite{Menard2017,Menard2019b}.

Pb substrates are not suitable for controllably manipulating the adatom positions by an STM tip. Atom-by-atom construction of chains of Fe adatoms by means of the STM tip was possible on Re \cite{Kim2018} and Ta surfaces \cite{Kamlapure2018}. STS measurements on the Re system reveal zero-bias features (but no peaks) at both ends of the wire which are interpreted as evidence for MZMs in Ref.\ \cite{Kim2018}. Compared to Pb, Rhenium has a considerably smaller superconducting gap. Measurements with a magnetic tip suggest that the Fe adatoms on Re order into a spin spiral, presumably as a result of the strong Dzyaloshinsky-Moriya interactions on a 5d transition metal substrate.

Magnetic adatoms on conventional superconductors have specific advantages for studying topological superconductivity. STM provides spatial information on MZMs and adatom structures can be constructed by tip manipulation on appropriate substrates. The latter might ultimately allow one to track the emergence of MZMs as the chain is built up adatom by adatom, or enable the implementation of more involved structures \cite{Li2016}. At the same time, the larger energy scales, while stabilizing the topological phase,  make it challenging to manipulate the MZMs, in particular for strongly coupled adatoms.

\subsection{Metallic surface states with strong spin-orbit coupling}

The spin-orbit coupling that is found in semiconductors is weaker than the record couplings found in metallic systems. In particular, Au has a strongly Rashba split surface state. This makes it an attractive system in the context of the search for topological superconductivity and MZMs \cite{Potter2012}, albeit with the difficulty of a high electron density that makes confinement to quantum wires challenging. In a recent experiment, gold wires were grown on top of a superconducting Vanadium substrate and covered by islands of EuS, a ferromagnetic insulator \cite{Manna2019}. It was found that the coupling to EuS drastically reduces the Fermi energy in the Au wires and consequently the number of transverse channels. Zero-bias peaks observed by STS at the edges of EuS islands covering the Au nanowires when applying a magnetic field were interpreted as (strongly localized) MZMs.

\section{Proving and disproving Majorana phenomenology}
\label{sec:prove_disprove}

The study of Majorana zero modes in condensed matter systems is strongly motivated by the prospect of topologically protected operations on Majorana qubits and the search for Majorana bound states has indeed stimulated enormous progress in the material science and device fabrication of hybrid systems. A significant body of transport experiments has provided data which are consistent with an interpretation in terms of Majorana bound states. But despite this success, the community remains unable to conclude whether single localized Majorana bound states have been observed.

The difficulty of demonstrating the existence of MZMs is inherent in two of their defining properties. First, MZMs are largely robust against many interactions with their environment. Given that experiments typically rely on the response of the measured system to perturbations introduced by the experimental apparatus, this is a difficulty. As described above, several properties of MZMs allow one to circumvent this challenge, most notably the tunneling conductance at zero bias. Second, identifying features of a MZM should be insensitive to variations of parameters. This is a difficulty as it forces one to look for negative results. Clearly, there are numerous parameters that would not affect a zero-bias peak in the conductance, but for most this is just reflecting a lack of coupling to the measured system. This difficulty can be handled by studying how parameters cease to affect one-dimensional topological superconductors as they become longer and the MZMs tend to decouple.

In addition to these two inherent difficulties, there is another challenge that obstructs the identification of MZMs in several of the experimental systems, namely, the possible existence of multiple almost-zero-energy end states which exist in parallel with (or instead of) the MZMs. These may have topological or nontopological origins. In particular, under certain conditions, both multi-subband wires and planar Josephson junctions approximately obey a composite anti-unitary symmetry that places them in symmetry class BDI. In contrast to the binary $\mathbb{Z}_2$ classification of class D, which allows for only one Majorana mode at each end, class BDI is characterized by a topological $\mathbb{Z}$ index which counts the number of Majorana zero modes at the system's ends and can take on any integer value (see Box 1).

The simplest example of this difference is furnished by wires with several subbands at the Fermi energy \cite{Kell2012,Liu2017,Vuik2019}. When superconductivity is sufficiently weak, only electrons with opposite momenta in the same subband are paired. Each subband may be regarded as a separate single-subband wire realizing either a trivial or a topological superconductor. The number of MZMs at the ends is then equal to the number of topological subbands. A variation of the electrostatic potential towards the wire's end may induce individual subbands to become trivial close to the wire's end. This pushes the MZMs of those subbands into the wire's bulk, thus changing the number of modes that are coupled to the leads. Moreover, the additional symmetry that protects the MZMs at the ends from hybridization is nonlocal and as such easily broken, e.g., by disorder or the confinement potential close to the system's ends. Then, MZMs on each end may hybridize and split, possibly leaving different numbers of unsplit modes on the two ends. In the simplest example, two modes may be left unsplit on one end, appearing as a zero-energy Andreev state \cite{Moore2018, Stanescu2019, Avila2019}, with no modes on the other end. In the presence of such a zero-energy  Andreev state, the system has the same degeneracy of the ground state as a ``proper'' one-dimensional topological superconductor, but the state is not protected, and may shift away from zero energy under the effect of a perturbation. Furthermore, such a state would behave differently from topologically defined MZM when attempting to braid.

In principle, the experimental tools we described so far should be able to distinguish between a one-dimensional topological superconductor with a single MZM on each end and a system with a different distribution of MZMs. In practice, this is a difficult task. Tunneling measurements, regardless of whether they are based on a scanning tunneling microscope or static tunnel contacts, measure tunneling peaks at zero voltage bias when they couple to MZMs. But they do not easily discern the number of MZMs. In the right range of temperatures, the height of the zero-bias peak caused by a single MZM is universal while the height of a a similar peak caused by several overlapping MZMs is nonuniversal. However, there may be other reasons for not observing the universal value. Coulomb-blockade experiments, in principle, probe whether the ground state is degenerate due to the presence of MZMs. However, they do that only if the MZMs couple to the outside world and, moreover, the experiments cannot easily discern between different possible numbers of degenerate ground states. Experiments probing the system by electromagnetic waves, measuring for example absorption or reflection, couple only to excitations that conserve the number of electrons and therefore do not probe ground state degeneracies. Photon-assisted tunneling in transport experiments are different and provide additional energy resolution \cite{Zanten2019,Peters2020}, but do not give fundamentally different information from dc transport experiments.

Indeed, features which have been taken as evidence of Majorana physics in transport experiment have also been explained in terms of model systems with an even number of spatially-overlapping Majorana bound, or so-called Andreev states \cite{Kell2012,Lee2012,Liu2012,Liu2017,Ptok2017,Vuik2019,Penaranda2018,Moore2018,Reeg2018,Setiawan2019,Junger2020} (for a recent review on transport spectroscopy of Andreev and Majorana states, see Ref.\ \cite{Prada2019}). Transport experiments which discern the multiple-Majorana from the true single-Majorana scenario by probing the coherence of the nonlocal fermion made out of two MBS \cite{Fu2010,Hell2018,Drukier2018} have been proposed, with some experimental progress in this direction \cite{Whiticar2019}.

With these difficulties in mind, it is plausible to expect that transport experiments of the types that have been performed so far will confirm the existence of MZMs only with an ever growing level of confidence. An improved set of transport experiments to accumulate evidence for the existence of a single pair of MZMs in a one-dimensional topological superconductor could combine measuring the tunneling density of states on both ends simultaneously with a measurement of the bulk energy gap, in such a way that the trivial phase, the topological phase, and the gapless transition points between them can all be identified and shown to be insensitive to gap-preserving variations of relevant parameters such as magnetic field or chemical potential (see \cite{Grivnin2019} for an initial attempt in this direction).

A more refined level of confidence will, however, most likely require new types of experiments that address the nonlocal and nonabelian properties of Majorana states (see Box 3 for an introduction to the nonabelian properties). An important class of experiments would probe the low-energy subspace of many-body states that is spanned by the MZMs. At fixed fermion parity and in the absence of above-gap Bogoliubov quasiparticles, a system with four MZMs spans a two-dimensional subspace and can be regarded as realizing a single qubit. Multiple qubit degrees of freedom can be realized in systems containing more than four MZMs \cite{NayakReview,OregOppenAnnualReview}. Variations of the fermion parity can be suppressed when placing the MZMs on Coulomb-blockaded islands. This has been demonstrated for semiconductor quantum wires, including the $S$-stripe geometry based on two-dimensional heterostructures, but is clearly less natural in other Majorana platforms. At sufficiently low temperatures, above-gap quasiparticles are suppressed by virtue of the superconducting gap. In practice, fermion parity will only be preserved for a finite duration and it would be important to measure the lifetime of a fixed parity state. Initial experiments of parity lifetimes, though not in the topological regime, have been performed and give encouraging results \cite{Higginbotham2015,Albrecht2017}. Within the fermion parity lifetime, the qubit degrees of freedom can be described in terms of Pauli operators composed of products of two MZMs each. Coupling judiciously to these two MZMs, one can in principle read out these Pauli operators. Various readout schemes have been discussed in the literature \cite{Hyart2013,Aasen2016,Plugge2017,Karzig2017,Schrade2018,Leijnse2012,Hoffman2016}, all utilizing that when splitting the ground state degeneracy by coupling two (or more) MZMs, the parity can be read out by charge, energy, or spin measurements.

This readout capability would open the door to probing the dynamics of the system within the qubit subspace. In an ideal topological qubit, the overlap of the MZMs is negligible and all qubit states have the same energy. Experiments can then probe the (anti)commutation relations between the Pauli operators of a qubit. Consecutive projective measurements of the qubit Pauli operators give characteristic results, as familiar from Stern-Gerlach experiments, and would provide strong evidence for the realization of MZMs. Alternatively, transport measurements probing interference can implement simultaneous weak measurements of two distinct Pauli operators and reveal signatures of their noncommutativity \cite{Manousakis2020}. Hybridizing the MZMs introduces additional (unitary) dynamics within the qubit subspace. Controlling the coupling between the MZMs, e.g., via an intermediate quantum dot, can thus be used to implement standard experiments probing the quantum dynamics of qubits. The noise-induced dephasing of this quantum dynamics of Majorana qubits may also be a sensitive probe of their nonlocal nature and help distinguish truly isolated MZMs from situations which are akin to Andreev bound states with near-zero energy \cite{Mishmash2020}.

It is important to stress that the above-mentioned experiments relying on the degenerate subspace spanned by MZMs critically depend on the topological state having a sufficiently hard gap. Otherwise, the quantum state stored in the parity degree of freedom will be corrupted by quasiparticle exchanges with non-Majorana subgap states, resulting in a very short parity lifetime. It is presently unclear whether a sufficiently hard gap can be realized in any of the engineered topological superconductors. A fundamental obstacle might be the absence of Anderson's theorem for $p$-wave superconductors. Thus, imperfections in the crystal structure or inhomogeneous electric or pairing potentials will in general induce subgap states. With an almost continuum of trivial (Andreev) subgap states, the parity degree of freedom associated with the MBSs effectively spreads out. Topological protection may still be maintained if the subgap states are localized by disorder, and the separation of the MZMs is larger than the localization length of the in-gap states \cite{Akhmerov2010}, but this condition makes it increasingly difficult to control and read out the Majorana joint parities \cite{Munk2019,Munk2020,Steiner2020}. Judging from the currently available experimental data, the gap is unfortunately rather soft on the topological side of the transition, constituting a major challenge for probing the nonlocal and nonabelian properties of Majorana states.

\section{Outlook}
\label{Sec:outlook}

Once it is experimentally verified that a set of MZMs spans a degenerate subspace of ground states with a sufficiently long parity lifetime, braiding experiments probing the nonabelian properties of MZMs come within reach. There are various proposals for experimental setups that could probe and exploit braiding of MZMs. A full account of these ideas is beyond the scope of this review (see Refs.\ \cite{BeenakkerBraidReview} and \cite{LutchynReview} for reviews which discuss this in more detail). Braiding can be implemented in a variety of ways (see also Box 3). The most direct implementation relies on actually braiding Majorana states in real space \cite{Alicea2011}. Alternatively, braiding can be performed in parameter space, using couplings to auxiliary Majoranas, either by controlling the couplings directly \cite{Sau2011} or by using Coulomb blockade to mediate interactions \cite{Hyart2013,Aasen2016}. Finally, measurement-based protocols \cite{Bonderson2008} can effectively emulate braiding processes of Majorana states \cite{Plugge2017,Karzig2017}.

MZMs spanning a degenerate subspace are also a prerequisite for quantum-computing applications of MZMs. There have been numerous proposals for implementing quantum information storage and quantum information processing using Majorana-based qubits \cite{Kitaev2003,Bravyi2006,Bonderson2008,NayakReview,Hyart2013,Ginossar2014,Barkeshli2015,Hoffman2016,Karzig2017,Plugge2017,Litinski2017,Litinski2018,OregOppenAnnualReview}. While MZM-based schemes do not allow for universal quantum computation in a fully topologically protected manner, quantum information storage and some (braiding-based) quantum gate operations can be topologically protected (see Box 3). The remaining gate operations required for universal quantum computation can in principle be implemented with high levels of accuracy \cite{Bravyi2005,Bravyi2006,Karzig2016,Karzig2017,Litinski2018,Karzig2019}. However, these schemes tend to involve projective measurements of Majorana qubit operators, emphasizing again the importance of implementing and understanding readout schemes for Majorana qubits. It is interesting to note that Majorana-based qubits have other unique features which may be useful for quantum information processing. In particular, the qubits can be read out in the Pauli bases to exponential precision in the distance of the MZMs. This is a consequence of the fact that the qubit's three Pauli matrices are composed of distinct products of two MZMs. Moreover, in some qubit implementations, readout is equally easy in the three Pauli bases and can be easily switched between them. This potentially allows for a significant reduction in the number of hardware-based gate operations by virtue of the Gottesman-Knill theorem \cite{Litinski2018}.

The nonabelian nature of MZMs in conjunction with Coulomb charging effects provide a starting point for numerous intriguing constructions (see also \cite{OregOppenAnnualReview} for a review). In the context of quantum computation, one can imagine to use networks of MZMs to implement topological error correcting codes, including the toric or surface code \cite{Vijay2015,Landau2016,Plugge2016,Litinski2018} as well as color codes \cite{Litinski2017}.
While it is hoped that MZM-based hardware is intrinsically error resilient, the corresponding qubit lifetimes will still be finite. It may then be attractive to combine MZM-based hardware with topological quantum error correction. The error resilience of MZM-based qubits would hopefully make it sufficient to work with small logical qubits, thus limiting the required hardware overhead. Very similar constructions based on Majorana networks can be used to implement various intriguing topological or strongly interacting models such as the Kitaev spin models \cite{Kells2014,Sagi2019,Thomson2018}, Sachdev-Ye-Kitaev models \cite{Alicea2017,Pikulin2017}, or realizations of supersymmetric models \cite{Ebisu2019}. Coupling the spin-like parity degrees of freedom to electronic leads is also associated with remarkable Kondo physics \cite{Beri2012,Altland2013}, which may be experimentally much easier to realize and constitutes an interesting test of Majorana physics. All of these exotic ideas rely on the phenomenology of nonlocal quantum information expected for Majorana states. Engineering systems with topologically protected ground-state degeneracies (or ruling out that it can be done) would thus have a substantial impact on future studies of these intriguing systems.

Readers who hope that this paper ends with the authors' bet for the date by which nonabelian physics will be harnessed for processing  topologically protected quantum information are heading for a disappointment, we are afraid. In our opinion, it does seem that the observed MZM signatures in the systems that we reviewed are caused by the relevant physics, \textit{i.e.},  roughly the right combination of spin-orbit interactions,  superconductivity, and breaking of time-reversal symmetry. However, it could be that there still need to be critical developments in the material and device-fabrication technology to reach a point where the nonabelian nature of the MZMs will dominate the physical properties of the systems, and will survive the detrimental effects of in-gap states and unwanted almost-zero modes.

\section{Box 1: Topological superconductivity and Majorana zero modes}
\label{Box1}

Topological phases of fermionic systems with a bulk gap are frequently characterized by gapless boundary modes. Historically, the first example of this bulk-boundary correspondence were the chiral edge modes of the integer quantum Hall states, later complemented by the helical edge modes of the quantum spin Hall state and many more. Topological superconductors also support gapless edge states, which are Majorana modes as a result of particle-hole symmetry. In one-dimensional systems, these edge modes are MZMs localized at the ends of the wire or at domain walls between the topological superconductor and a gapped nontopological phase.

In superconductors, fermionic quasiparticles are superpositions of electrons and holes. Their effective lack of charge conservation is made possible by the superconducting condensate of Cooper pairs with charge $2e$. The creation operator of a quasiparticle of energy $E$ has the form
\begin{equation}\label{MZMoperator1}
\begin{aligned}
    \gamma_{E}&=&\int \mathrm{d}x \big[f^{{}}_{E,h,\uparrow}(x) \Psi^{{}}_\uparrow(x)+f^{{}}_{E,h,\downarrow}(x) \Psi^{{}}_\downarrow(x)\\
    && +  f_{E,e,\uparrow}^*(x) \Psi^\dag_\uparrow(x)+f^*_{E,e,\downarrow}(x) \Psi^\dag_\downarrow(x)\big],
\end{aligned}
\end{equation}
where $\Psi^\dag_\sigma(x)$ is the electron creation operator that creates an electron with spin $\sigma$ at position $x$, and
the subscripts $h,e$ stand for holes and electrons. The particle-hole symmetry inherent to superconductors implies that each quasiparticle described by Eq.\ (\ref{MZMoperator1}) has a partner of energy $-E$ for which electron and hole amplitudes are exchanged and complex conjugated:
\begin{eqnarray}
f_{-E,h,\sigma}=f^*_{E,e,\sigma}\\
f_{-E,e,\sigma}=f^*_{E,h, \sigma}
\label{phsymmetry}
\end{eqnarray}
A localized zero energy mode ($E=0$) therefore satisfies $f_{e,\sigma}=f^*_{h,\sigma}$ for both spin components. Thus, for a MZM the amplitudes of electrons and holes have the same absolute values.  Any deviation from $E=0$ tips the balance towards one sign of charge.

Furthermore, Eq.\ (\ref{phsymmetry}) implies that for a MZM, the quasiparticle is self-adjoint, $\gamma_{E=0}=\gamma^\dag_{E=0}$.
Therefore, $\gamma_{E=0}^2$ does not vanish, there is no Pauli principle, and no notion of  a single Majorana zero mode being full or empty. This familiar fermionic intuition comes back when considering two Majorana zero modes, $\gamma_1$ and $\gamma_2$. These can be combined into an ordinary (complex) fermion mode $d=\frac{1}{2}(\gamma_1 + i\gamma_2)$, which can be empty or occupied, respecting the Pauli principle. As the Majoranas making up this complex fermion are gapless, the energy of the system does not depend on the occupation $d^\dagger d$. Consequently, each pair of Majorana zero modes implies a twofold degeneracy of the ground state of the topological superconductor. As mentioned, the existence of Majorana zero modes is protected and therefore this degeneracy is also topologically protected and exact, up to corrections arising from the spatial overlap between the MZMs. In schemes for a Majorana-based topological quantum computer, the manifold of degenerate ground states is exploited to host the qubits.

We can now  understand the resilience of the system's ground states to local noise: A local (for example electrical) disturbance couples to the local charge in the superconductor: $\rho(x)=\Psi^\dag_\uparrow(x)\Psi^{{}}_\uparrow(x)+\Psi^\dag_\downarrow(x)\Psi^{{}}_\downarrow(x)$. If the extent of the disturbance is smaller than the distance between the Majorana end modes, there is no way that it can involve both $\gamma_1$ and $\gamma_2$. The quantum information stored in the $d$ fermion degree of freedom is hence immune to (local) noise, at least up to corrections controlled by the overlap of the localized MZMs. Their localization length, $\xi = \hbar v_F/\Delta$, is determined by the bulk gap, $\Delta$, of the topological superconductor and the Fermi velocity, $v_F$. Thus, in principle the  overlap between MZMs can be made exponentially small when their separation is large compared to the coherence length $\xi$.

A systematic approach shows that there are different topological superconducting phases, depending on the symmetries of the system \cite{Ryu2010}. The generic case (referred to as class D) has only particle-hole symmetry. It has a single MZM localized at each end of the one-dimensional system in its topological phase. In the presence of further symmetries, there may be more than one MZM at each end of a one-dimensional system, even when they spatially overlap. Some one-dimensional models exhibit a chiral symmetry (class BDI) and have arbitrary numbers of MZMs localized at their ends. Time-reversal-invariant topological superconductors are characterized in 1D by a $\mathbb{Z}_2$ index reflecting the presence or absence of a Kramers {\it pair} of MZMs at each end. Generally, a multitude of MZMs at each end is an obstacle for quantum information processing, since the symmetry that protects the degeneracy of the MZMs when they are static does not protect their decoupling when they are braided \cite{Wolms2016}.

\section{Box 2: Spectroscopy of Majorana zero modes}

\begin{figure*}
\centerline{\includegraphics[width=1.\textwidth]{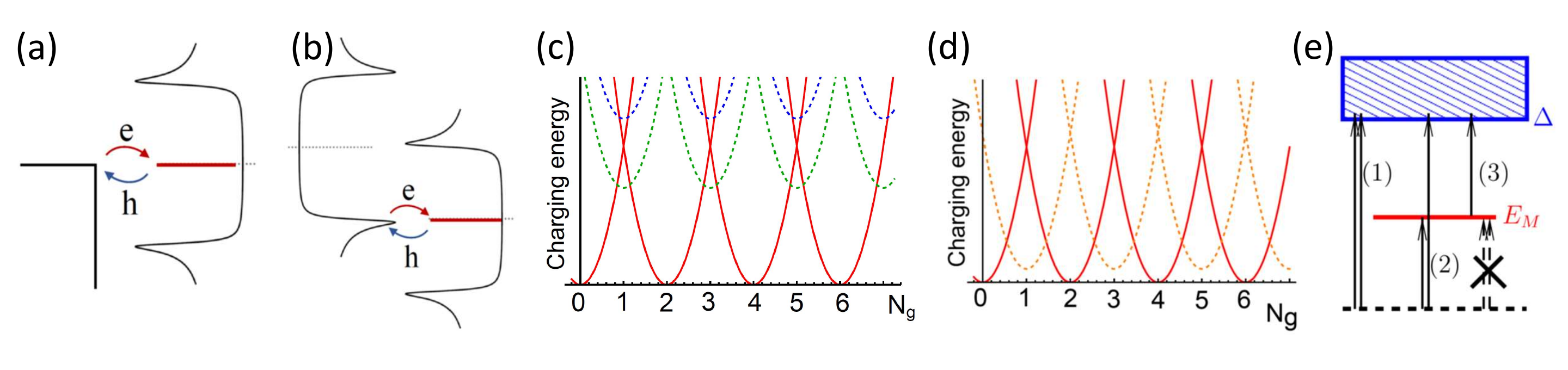}}
\caption{\label{fig:spectroscopy}
Illustration of selected spectroscopic techniques. (a) Resonant Andreev reflection process probing a MZM with a normal-metal electrode. (b) Same for a superconducting electrode. (c) Charging energy of a Coulomb-blockaded wire in the topologically trivial phase. The charging parabolas for odd electron numbers (blue dashed lines: $\Delta>E_C$ with electron-pair additions; green dashed line: $\Delta<E_C$ with single-electron additions) are shifted upward relative to the parabolas for even electron numbers (red) by the pairing energy $\Delta$. (d) Charging energy of a Coulomb-blockaded wire in the topological phase. The charging parabolas for odd electron numbers (dashed orange) are shifted upward by a small hybridization energy of the MZMs, which decreases to zero with increasing distance between the MZMs. (e) Excitation of quasiparticles by a microwave field. A photon breaks a Cooper pair into two above-gap quasiparticles (process 1) or into the hybridized Majorana mode at energy $E_M$  and an above-gap quasiparticle (process 2). If the hybridized Majorana mode is occupied, the subgap quasiparticle can also be excited into the above-gap continuum (process 3). A process in which both electrons of the Cooper pair are excited into the hybridized Majorana mode is impossible, since the latter is nondegenerate. Panel (e) reprinted from \cite{Peng2016}.
}
\end{figure*}

Experiments recording current-voltage characteristics frequently probe the existence of subgap states at zero or near-zero energy. When tunneling into a topological superconducting wire through a tunnel barrier from a normal-state electrode, the zero-energy end state will lead to a zero-bias peak in the differential conductance. At subgap voltages, tunneling occurs via resonant Andreev reflection with an electron tunneling, say, into the wire and a hole tunneling out into the lead, see Fig.\ \ref{fig:spectroscopy}(a). Both electron and hole have tunneling amplitudes of equal magnitudes as a consequence of particle-hole symmetry. This implies that at zero temperature, the zero-bias peak is quantized to $2e^2/h$ \cite{Sengupta2001,Law2009,Flensberg2010} and has a width given by the coupling strength between the Majorana bound state and the lead. As temperature becomes larger than the zero-temperature width, the peak reduces in height and broadens.

Majorana zero modes have also been probed with superconducting leads, especially in the context of STM experiments \cite{NadjPerge2014,Ruby2015,Feldman2017}. The superconducting gap $\Delta$ suppresses thermal excitations in the lead and the sharp BCS singularity in the density of states provides improved resolution. The basic tunneling process is the same as for normal-metal leads, except that electron and hole in the lead now require a minimal excitation energy of $\Delta$, see Fig.\ \ref{fig:spectroscopy}(b). As a result, the Majorana peak will appear at bias voltages of $eV=\pm \Delta$. The peak height is no longer quantized but depends on the superconducting density of states of the lead (STM tip). For a conventional BCS density of states, one finds a value of $(4-\pi)2e^2/h$. Moreover, particle-hole symmetry implies that the peaks at positive and negative bias voltages are symmetric \cite{Peng2015}.

When a semiconducting quantum wire is proximity-coupled to  a superconductor which is sufficiently small and is only weakly connected to the outside world, the hybrid semiconductor-superconductor wire has a finite charging energy $E_C$. When changing a gate voltage applied to a wire in a nontopological superconducting state, electrons are added in pairs ($2e$ processes) when $\Delta > E_C$ and individually ($1e$ processes) when $E_C>\Delta$. In the latter case, the gate voltage spacing between consecutive additions of electrons exhibits an even-odd alternation, reflecting the fact that an odd electron remains unpaired and requires an additional energy cost of $\Delta$, see Fig.\ \ref{fig:spectroscopy}(c). For a topological superconducting wire, this extra energy cost does not exist since the odd electron can occupy the zero-energy fermion associated with the Majorana end states [Fig.\ \ref{fig:spectroscopy}(d)]. Experimentally, this phenomenon is reflected in the gate-voltage spacing between Coulomb blockade peaks, and one observes characteristic changes in these spacings when driving a hybrid semiconductor-superconductor wire from the trivial into the topological phase by an applied magnetic field \cite{Albrecht2016}.

Coupling two superconductors by a tunnel junction or a normal-metal region leads to the Josephson effects. When coupling two conventional superconductors, Cooper pairs moving between the superconductors lead to a Josephson current which is periodic in the phase difference $\phi$ of the two superconducors with period $2\pi$. When coupling two topological superconducting wires, the pair of MZMs hybridize across the junction, allowing for the coherent exchange of single electrons which manifests itself in a $4\pi$-periodic Josephson effect \cite{Fu2008b,Lutchyn2010,Oreg2010}. Due to hybridization of the MZMs, the ground-state energy varies as $\pm E_M \cos(\phi/2)$. The sign depends on the fermion parity of the junction, so that the $4\pi$-periodic Josephson effect can only be observed in the absence of quasiparticle poisoning processes, such that the parity does not fluctuate. While this may be hard to satisfy in $dc$ experiments, signatures of topological superconductivity may  persist in the presence of quasiparticle poisoning when measuring on sufficiently short time scales, for instance by measuring Shapiro steps \cite{Jiang2011,Dominguez2012,SanJose2012,Houzet2013}, noise correlations \cite{Houzet2013}, or Josephson radiation \cite{Deacon2017}.

The topological nature of Josephson junctions can also be revealed by spectroscopically probing the subgap states localized at the junction.
While tunneling changes the fermion number parity, microwave experiments give access to excitations at fixed fermion parity. Thus, at zero temperature, absorption of a microwave photon must be associated with the creation of two quasiparticles \cite{Bretheau2013,Bretheau2013b,Virtanen2013,Vayrynen2015,Peng2016}.
In the absence of subgap states, the minimal energy cost is given by $2\Delta$. For a short topological Josephson junction, one of the quasiparticles can occupy the subgap state formed by the pair of MZMs, which allows one to measure the characteristic phase dependence of its energy in suitable setups. The phase-dependence of the subgap spectrum can also be probed by other techniques such as switching current measurements \cite{Zgirski2011,Peng2016}.

\section{Box 3: Nonabelian properties and topological quantum computing}

Perhaps the most coveted property of Majorana zero modes is their nonabelian exchange properties, sometimes referred to as ``nonabelian statistics" even though the MZMs are not particles. As discussed in Box 1, the ground state becomes $2^N$-fold degenerate in the presence of $2N$ MZMs. Adiabatically exchanging the positions of the Majoranas leaves the system in this ground-state manifold, but not necessarily in the same state. Instead, the state changes by a unitary transformation, which depends on the order in which the exchanges are being performed. This nonabelian exchange ``statistics"  contrasts sharply with the case of bosons, fermions, or abelian anyons, for which exchanges merely modify the state by factors of $+1$, $-1$, or $e^{i\alpha}$, respectively.

Physically, braiding operations can be implemented most directly by explicitly moving Majoranas around each other in real space. This can be done for Majoranas in two dimensions or on a network of wires \cite{Alicea2011}. In either case, the location of the MZM is tied to a defect such as a vortex or a domain wall whose location can be manipulated. For a defect trajectory to correspond to a nontrivial braiding operation, the ground-state degeneracy has to remain unchanged throughout the trajectory, but must change somewhere along the way when the trajectory is contracted to a point. Such braiding operations in real space are a particular case of a more general notion. Braiding operations satisfying the above conditions can also be implemented in other parameter spaces, for instance by judiciously varying overlaps between pairs of MZMs \cite{Sau2011,Hyart2013,Aasen2016}.

The unitary transformation implemented by braiding $\gamma_i$ with $\gamma_j$ may be obtained from rather general arguments. Since the MZMs are hermitian, we expect $U\gamma_iU^\dagger=\pm \gamma_j$, where the sign remains to be determined. Moreover, $U$ should conserve fermion number parity and involve only the braided MZMs, $\gamma_i$ and $\gamma_j$. These considerations fix $U$ to be
\begin{equation}
   U_{ij} = \frac{1}{\sqrt{2}} (1 + \gamma_i\gamma_j).
\end{equation}
One readily checks that these transformations do not necessarily commute.

Experimentally, braiding can be implemented by using a variety of knobs -- such as gate voltages, charging energies, or supercurrents -- that move MZMs around and/or control their couplings. The simplest braiding scheme involves two wires, with the pair of MZMs of each wire initialized in a state with, say, even fermion number parity. A single braid of MZMs from the two wires would transform the system into a superposition of even and odd parities, with equal probabilities. Strikingly, two consecutive braids would leave both wires in odd-fermion-number states. This outcome is independent of the details of the braiding process as long as the ground-state degeneracy remains unchanged.

The nonlocal and nonabelian properties are at the heart of quantum information processing in topological quantum computers. Information stored in the manifold of ground states is topologically protected against local sources of decoherence as long as the excitation gap does not close. However, this also implies that local (say, electric or magnetic) fields do not manipulate the quantum information. Here, the nonabelianess comes to the rescue. Braiding processes can be used to change the state of the qubits and thus to process the quantum information, depending only on the topology of the exchange process and not on the detailed trajectory. In general, universal quantum computation can be realized on the basis of the single-qubit Clifford gates (Hadamard and phase gate), the $T$-gate, and the entangling controlled-NOT (CNOT) gate acting on a pair of qubits. For a Majorana-based quantum computer, braiding alone only implements the single-qubit Clifford gates. It thus cannot implement universal quantum computation in a manner which is fully topologically protected. Nevertheless, the CNOT gate can be implemented when including additional ancilla qubits combined with two-qubit-parity measurements.
The $T$-gate can in principle be enacted in an accurate, though not topologically protected manner, by magic state distillation \cite{Bravyi2006}. Interestingly, schemes for universal Majorana-based quantum computation need not even involve any explicit braiding operations when performing the single-qubit Clifford gates by measurements \cite{Bonderson2008} or when effectively outsourcing the corresponding operations to a classical computer \cite{Karzig2017,Litinski2018}.

\acknowledgments

We acknowledge support from ERC projects LEGOTOP (AS) and NONLOCAL (KF), Israel Science Foundation within the  ISF-Quantum program (AS),  Deutsche Forschungsgemeinschaft through CRC 183 (AS, KF, FvO), Danish National Research Foundation (KF),  Independent Research Fund Denmark | Natural Sciences (KF), and QuantERA project Topoquant (FvO).

%

\end{document}